\documentclass[showpacs,10pt,preprint,final,twocolumn,aps]{revtex4}%
\usepackage{amsfonts}
\usepackage{amsmath}
\usepackage{amssymb}
\usepackage{graphicx}%
\begin{document}
\input epsf.sty \flushbottom

\title{Are current-induced forces conservative?}
\author{M. Di Ventra\cite{MD}}
\affiliation{Department of Physics, Virginia Polytechnic Institute and State University,
Blacksburg, Virginia 24061}
\author{Yu-Chang Chen}
\affiliation{Department of Physics, Virginia Polytechnic Institute and State University,
Blacksburg, Virginia 24061}
\author{T.N. Todorov}
\affiliation{School of Mathematics and Physics, Queen's University of Belfast,
Belfast BT7 1NN, United Kingdom}
\pacs{73.40.Jn, 73.40.Cg, 73.40.Gk, 85.65.+h}

\begin{abstract}
The expression for the force on an ion in the presence of 
current can be derived from first principles without any assumption 
about its conservative character. However, energy 
functionals have been constructed that indicate that this force can 
be written as the derivative of a potential 
function. On the other hand, there exist compelling specific arguments 
that strongly suggest the contrary. We propose physical mechanisms that 
invalidate such arguments and 
demonstrate their existence with first-principles calculations. 
While our results do not constitute a formal resolution to the fundamental 
question of whether current-induced forces are conservative, they represent a 
substantial step forward in this direction. 
\end{abstract}

\maketitle

Electrons flowing in a conductor can transfer momentum to the ions.
From a static point of view, the steady-state electron density in a
current-carrying conductor is microscopically different from that 
in the absence of current.
In either picture, the passage of current 
generates additional forces on ions. This may result
in atomistic rearrangements and diffusion of atoms, a phenomenon known 
as electromigration.~\cite{sorbe} The advent of microelectronics has sparked off
a considerable interest in 
electromigration since it has been 
found to be a major failure mechanism in metallic interconnects.~\cite{blech} 
In recent years, a new type of electronics is being considered which envisions the use of 
nanoscale components. Since nanoscale conductors can carry current densities much
larger than their macroscopic counterparts, the problem of electromigration is again of
key importance. Recent work has studied the effect of current-induced forces in 
atomic~\cite{lango,agr,ita,yas,tod1,mioyang,mads} and molecular 
wires~\cite{aya,md1}. 
Combined with early investigations on the subject,~\cite{locdipole} 
this work has shed new light on the microscopic mechanism 
of current-induced mechanical effects. 

While a consensus on the 
physical origin of current-induced forces seems to have been reached, a fundamental 
question remains unanswered: are current-induced forces conservative? 
The difficulty in answering 
this question is intimately related to the difficulty in defining a force in a non-equilibrium 
problem.~\cite{md2,tod00}

In this paper we address this question in the following way: 
we consider representative arguments against 
the conservative nature of current-induced forces and suggest physical mechanisms 
that invalidate these arguments. We then corroborate the existence of 
such mechanisms with first-principles calculations. 
While this by no means {\it proves} that these 
forces are conservative, we believe 
it is a substantial step forward in this direction and hope that our 
suggestions will generate much-needed discussions.~\cite{prec} 

Let us start from the expression for forces in the presence of current. 
This expression can be derived 
in different ways. One approach is to start from a fully dynamical quantum 
description of the many-body problem 
comprising electrons and ions, and then take the 
limit as the ionic motion becomes classical.~\cite{md2} In this derivation, the
force expression comes about as the classical limit of Ehrenfest's
theorem applied to the rate of change of the momentum of an ion.~\cite{md2} 
Formally, this limit corresponds to high ionic masses or high 
kinetic energies \cite{tod01}. 
A different approach is to start from an appropriate mixed quantum-classical Lagrangian.
The force expression then results from the Euler-Lagrange equation of motion
for the classical ions \cite{tod01}. Both approaches result in the same
expression for the force 
on a given ion, with position ${\bf R}$, due to the self-consistent 
electronic density $\rho({\bf r})$ 
under current flow ~\cite{tod01}, 
\begin{equation}
{\bf F}= - \int d{\bf r}
\frac{\partial v}{\partial {\bf R}} \rho({\bf r}),
\label{equation}
\end{equation}
which can be equivalently written as~\cite{md2}
\begin{equation}
{\bf F}= -\sum_i \langle\psi_i\left|\frac{\partial v}{\partial
{\bf R}}\right|\psi_i\rangle -
\lim_{{\Delta}\rightarrow 0} \int_{\sigma}dE
\langle\psi_{\Delta}\left|\frac{\partial v}{\partial {\bf R}}\right|
\psi_{\Delta}\rangle.
\label{equation1}
\end{equation}
where $v$ is the electron-ion interaction.~\cite{comment}
The first term on the RHS of Eq.~(\ref{equation1}) is similar to the usual Hellmann-Feynman 
contribution to the force 
due to localized electronic states $|\psi_i\rangle$. The second term is the contribution 
to the force due to continuum states.~\cite{md2} The wavefunctions 
$|\psi_{\Delta}\rangle$ are eigendifferentials in the 
continuum $\sigma$.~\cite{md2} 
Due to the algebraic equivalence of the above two equations, in the following, 
when discussing current-induced forces, we will refer to either one without distinction. 

Let us now turn to the question of whether or not
these forces are conservative. It is essential 
to emphasize that the force obtained either from the 
Ehrenfest's theorem~\cite{md2} or a Lagrangian formulation~\cite{tod01}, is 
{\it not necessarily conservative} because the above first-principles 
dynamical derivations do not contain any assumption about the existence, 
or otherwise, of a free energy functional, of which this force might be the derivative. 
However, as a separate exercise, 
it was suggested both in Ref.~\onlinecite{md2} and in Ref.~\onlinecite{tod00} 
that forces in the presence of current can also be written as derivatives of 
appropriate energy functionals. 
For instance, in 
Ref.~\onlinecite{md2} the stationarity of an action was used to map 
the force problem into a variational one in which 
the force can be written as the derivative of an energy function. In 
Ref.~\onlinecite{tod00} an entropy function was used, whose maximization (constrained to the 
energy of the current-carrying system and the numbers of electrons in it
that originate from the right and left reservoirs) again
yields the force as the derivative of a potential function. 
These independent derivations suggest that current-induced forces
are conservative.

\begin{figure}
\includegraphics[width=.40\textwidth]{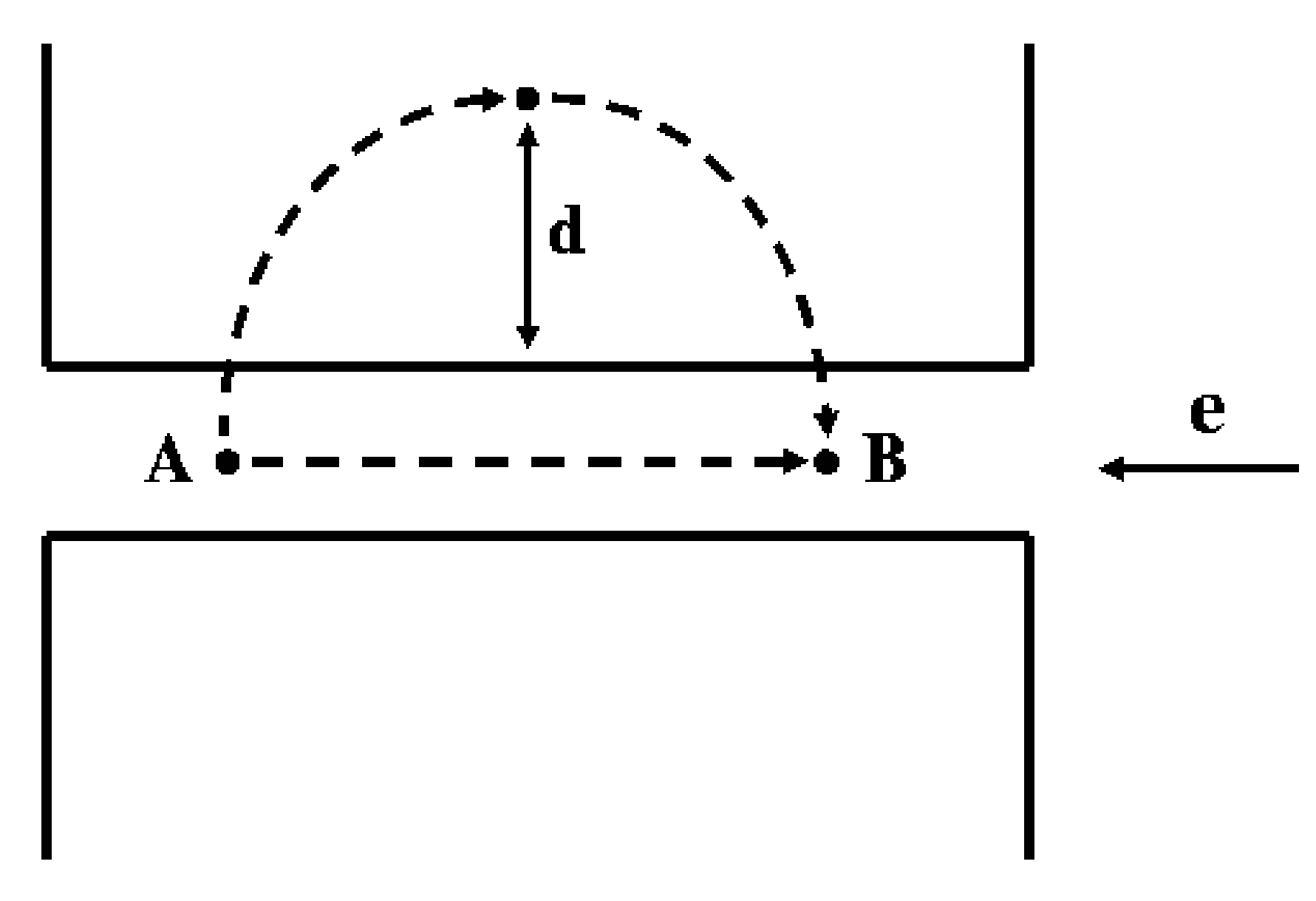}
\caption{Schematic of a current-carrying wire between two bulk electrodes. 
An atom is moved from a point $A$ to a point 
$B$ along two different paths indicated as solid lines. 
$d$ is the largest distance of the atom 
from the wire along the vacuum path. 
Electrons flow from right to left.}
\label{fig1}
\end{figure}

On the other hand, there exist compelling specific arguments that strongly
suggest that current-induced forces cannot be conservative. These arguments consist
in {\it gedanken} experiments, aimed at showing that the work
done by current-induced forces along specific closed
paths is not zero and that, therefore, these forces are not conservative. 
Below we refer to one notable such construction due to Sorbello.~\cite{sorbe} 
This example is of charming simplicity and is particularly convincing. The 
argument goes like this. Picture a single defect atom in a current-carrying wire~\cite{prec1} 
between bulk electrodes under an applied bias $V$. 
First, let the atom move, quasi statically, from a point $A$ to a point $B$ inside the wire
(see schematic in Fig.~\ref{fig1}).
The current-induced force on the defect in the wire does some non-zero
work on the way.
Now imagine taking the atom from $A$ to $B$ along a path 
that goes {\it outside} the wire (see Fig.~\ref{fig1}). 
Since outside the wire there apparently is no current, no current-induced force should be 
present. Hence, the work done by current-induced forces on the atom along the 
path from $A$ to $B$ inside the wire
cannot be the same as along the path outside the wire. This would 
be sufficient to prove that current-induced forces are not conservative.

However, there is something 
missing in this argument: the quantum-mechanical nature of current {\it outside} the wire. 
To see this, picture the atom at a point in space outside the wire. Let $|\psi\rangle$ be 
the global current-carrying electron states for the system. 
If there always is {\it some} (however small) 
quantum-mechanical coupling between the atom and the wire, there will
always be some non-zero tunneling probability for electrons to get from the wire
onto the atom, and back. The atom then
acts as a {\it resonant center} and there will always exist an energy window 
where the states $|\psi\rangle$ contain {\it resonant paths} that emerge from
the wire, pass through the atom, and go back into the wire. 
Due to the scattering of electrons at the atom, 
these resonant paths may be expected to create a charge dipole, centered at the atom, 
much like the local resistivity dipole in electromigration.~\cite{sorbe} 
Increasing the distance $d$ between the atom and the wire will make the resonant 
energy window narrower, while the resonance will remain pinned against the 
energy window $eV$, due to self-consistency. Now, the total current carried by the
resonance is given by the energy integral, over the energy window $eV$, 
of a transmission function, $T(E)$.
As the resonance gets narrower in energy, so does $T(E)$. 
$T(E)$ is always bounded and hence, 
as $d$ increases, the total current carried by the resonance will decrease. 
On the other hand, the charge dipole across the 
resonant center originates from the density of states, ${\rm DOS}(E)$, 
due to the populated current-carrying electron states in the energy window $eV$. 
By contrast with $T(E)$, ${\rm DOS}(E)$ is {\it unbounded}: if $d$ increases,
the resonant peak in ${\rm DOS}(E)$ gets narrower but taller in such a way that 
its energy integral remains constant. 
Since current-induced forces are related to the integral of the density of states 
[see e.g. Eq.~(\ref{equation1})], we expect the force on the atom 
due to the resonant dipole to be constant with $d$. 
In other words, there is always a current-induced force on the 
atom, even outside the wire, and this force does not vanish with increasing $d$.

A key assumption of our reasoning is that there {\it always} is {\it some} 
coupling between the atom in the vacuum and the metal.
If this coupling is truncated artificially, 
then the current and the ``resonant-dipole force'' 
will both be discontinuously extinguished. However, such a truncation
would at the same time create a bound state (at the atom) in the continuum, 
and would thus modify the nature of the electronic spectrum and of the corresponding 
Born-Oppenheimer (BO) manifold. Such a discontinuous modification 
corresponds to a breakdown of the BO approximation, in which case  
the notion of conservative forces is lost by definition. 

\begin{figure}
\includegraphics[width=.40\textwidth]{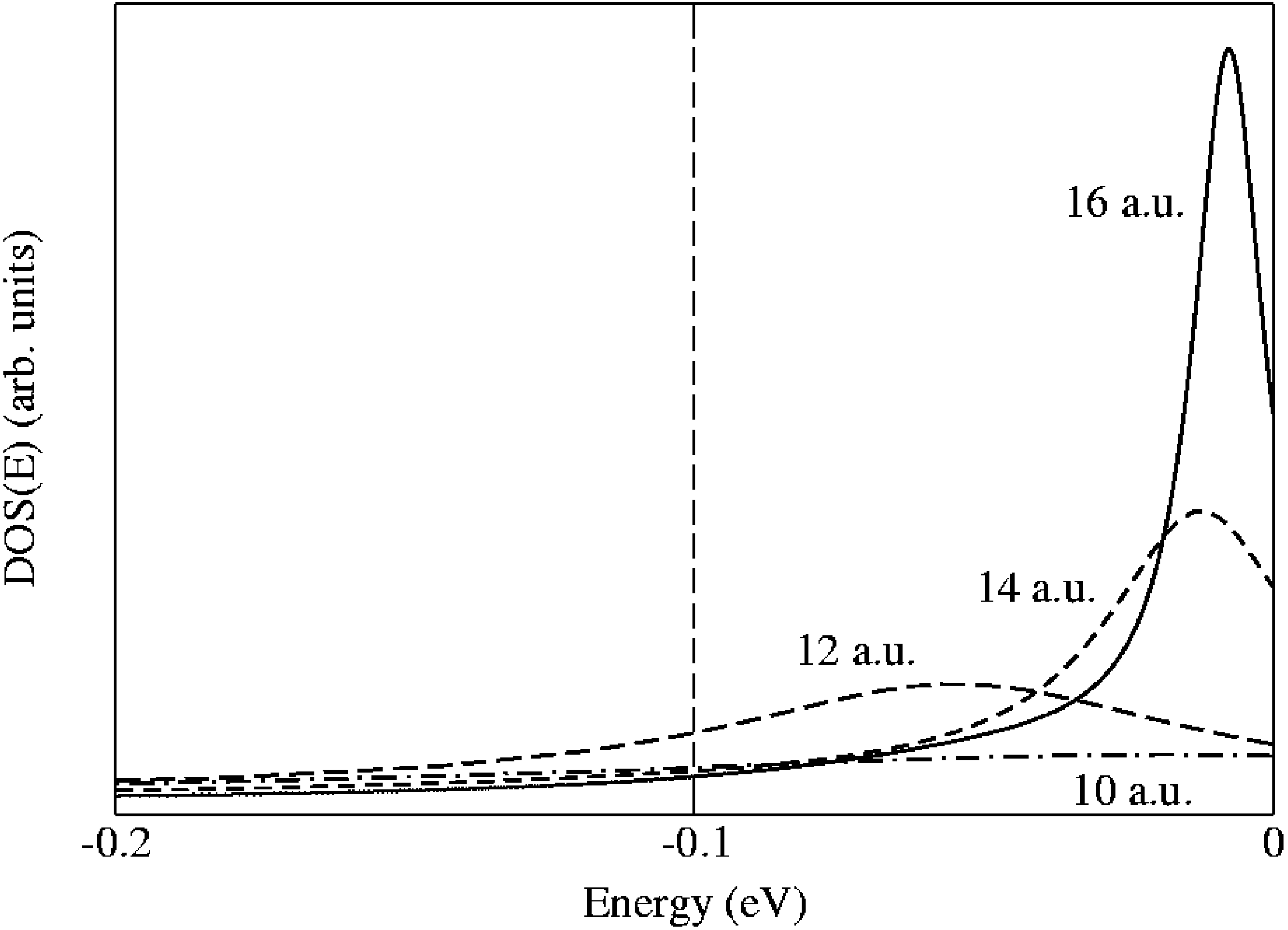}
\caption{Density of states due to the populated global current-carrying 
states originating from the right electrode for the system with the
Si atom, minus the same quantity for the electrode-electrode system without the atom,
at 0.1 V and for different separations of the jellium edges. Zero of energy corresponds to 
the right electrochemical potential. The vertical dotted line 
indicates the position of the left electrochemical potential.}
\label{fig2}
\end{figure}

Our analysis above relies only on generic properties of resonant
transmission. We therefore 
now seek to demonstrate the existence of a resonant dipole 
in the simplest possible resonant configuration: that of
a single atom in the vacuum between two bulk electrodes. Our goal is
to show that this resonant dipole gives rise to a finite current-induced
force that does not vanish with 
increasing electrode separation, even when the current dies out. 
To this end, we have performed first-principles 
calculations of the current-induced force on a
single Si atom between two jellium electrodes.~\cite{prec12} 
The interior electron density of the 
electrodes is taken equal to the value for metallic gold ($r_s\approx 3$).~\cite{prec13} 
The electrodes are kept at a certain bias with the left electrode positively biased so that 
electrons flow from right to left. In Fig.~\ref{fig2} we plot the density 
of states as a function of energy for a bias of 0.1 V and 
different distances between the jellium edges of the two electrodes. 
It is clear from this figure that a resonance develops in the energy window between the left and right 
chemical potentials. The resonance becomes 
narrower and taller with increasing distance. 
It is also evident from Fig.~\ref{fig2} that this resonance is pinned towards the 
right chemical potential. Such pinning can be intuitively understood as follows. 
By increasing the distance 
between the electrodes the resonance becomes narrower and narrower. 
In order to preserve charge neutrality, 
the amount of charge accommodated by the resonance needs to be the same at all distances. 
The only way to achieve this is for the 
``center'' of the resonance to move closer to the right electrochemical potential. 

\begin{figure}
\includegraphics[width=.40\textwidth]{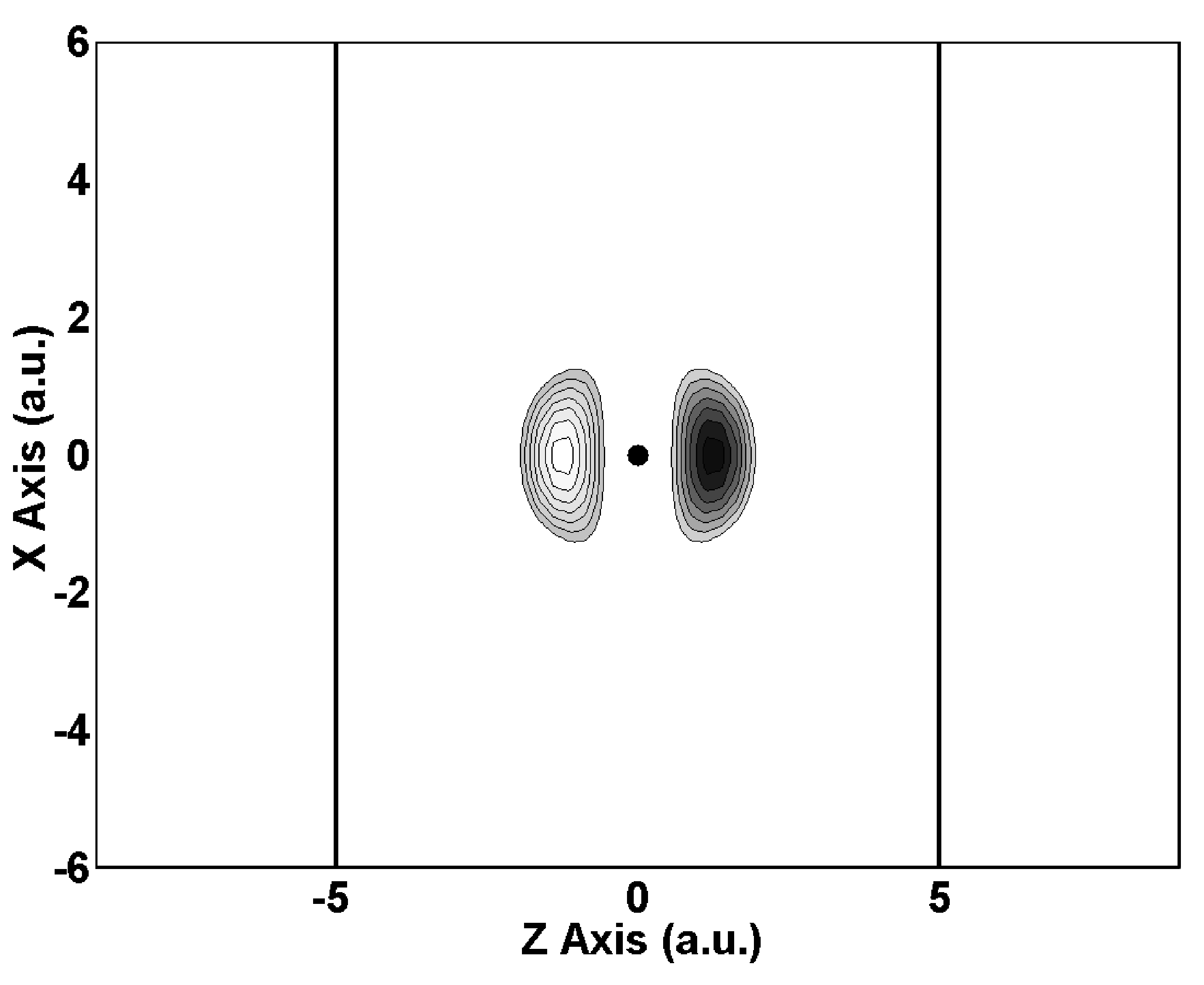}
\caption{Current-induced resonant vacuum dipole as defined in the text for a 
jellium-jellium edge distance of 10 a.u. Dark area on the right of the nucleus (represented 
by a dot) corresponds to negative charge.}
\label{fig3}
\end{figure}

Now that we have shown the existence of the resonance, let us look at the spatial 
distribution of the charge dipole it creates across the atom. 
In order to do this we follow Lang's approach~\cite{lango} and calculate 
first the difference between the electronic distribution at a given bias $\rho({\bf r},V)$ 
and the corresponding distribution at zero bias $\rho({\bf r},0)$. 
Call this quantity $\tilde{\rho}({\bf r},V)$. It contains three  
contributions: (i) the polarization of the biased electrode-electrode 
system without the atom, (ii) the polarization of the electron cloud at the atom 
due to the field generated by (i), 
and (iii) the current-induced resonant dipole. 
The first two terms generate a force of equal magnitude but 
opposite sign so they do not contribute to the total force. (An isolated atom 
in a uniform electric field does not have a net force on its nucleus). 
To eliminate (i) and (ii), we subtract (i) from 
$\tilde{\rho}({\bf r},V)$ and take the difference between the result of that subtraction
at a given electrode-electrode distance and 
the same quantity at a larger separation scaled to the 
field of the previous distance.~\cite{lango} This gives a negative fraction
of the current-induced dipole, so a change of sign is necessary. 
Call this quantity $\rho_f({\bf r},V)$. Finally, to 
isolate the antisymmetric (dipole) component of this distribution we evaluate 
$[\rho_f({\bf r},V) - \rho_f({\bf r},-V)]/2$. 
The result is plotted in Fig.~\ref{fig3} for a separation of 10 a.u. 
(with respect to 16 a.u.) and clearly shows the current-induced resonant 
dipole across the atom. 
The dipole has positive charge on the left of the 
atom and negative charge on its right 
so that the current-induced force pushes the atom to the right, i.e. against the electron flow. 

\begin{figure}
\includegraphics[width=.40\textwidth]{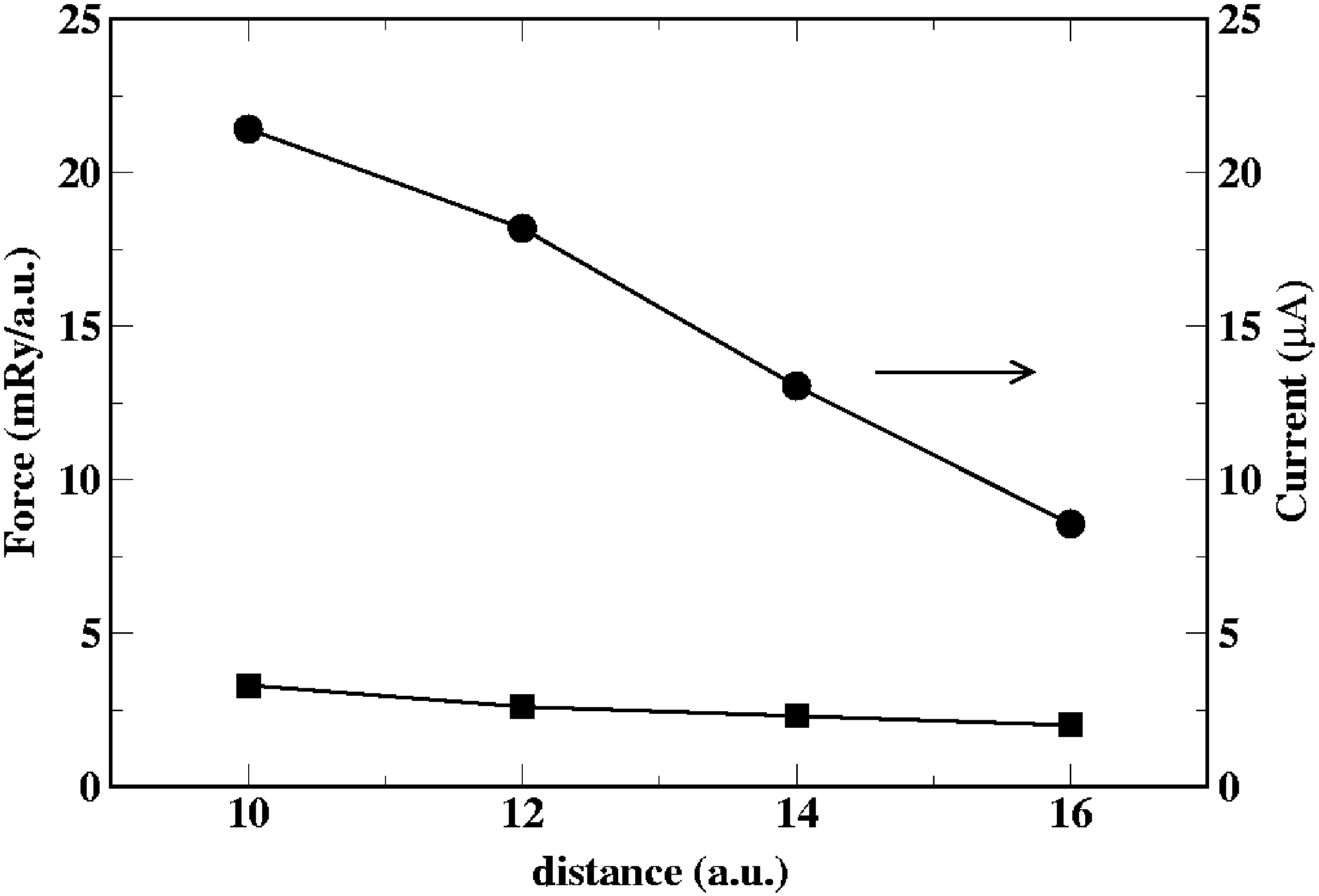}
\caption{Current (right scale) and current-induced force (left scale) for a 
single Si atom between two bulk electrodes as a function of 
distance between them.}
\label{fig4}
\end{figure}

Finally, we have to show that the current decreases with electrode 
separation while the force does not. This is plotted in Fig.~\ref{fig4} 
for different electrode separations. It is clear from Fig.~\ref{fig4} that 
the current decreases substantially with electrode separation 
while the force saturates at a value of about 2 mRy/a.u. 
Notice again that this force is positive, 
i.e. it pushes the atom against the electron flow, in agreement with 
the sign of the resonant dipole. These numerical results confirm our conjecture: 
as long as the atom is coupled to the electrodes, a resonant dipole forms across it, such that a 
non-zero current-induced force is exerted on the atom that persists with separation
even as the current dies out. 

We conclude by stressing again that our goal here is to show that 
existing arguments against the conservativeness 
of current-induced forces ignore specific physical mechanisms. These mechanisms
predict the existence of current-induced forces in unexpected 
places, such as an atom in the vacuum next to a current-carrying metal.
We have demonstrated the existence of such 
forces using first-principles calculations. These results do not 
{\it prove} that current-induced forces are 
conservative but we believe that they give extra support in favor of such a conclusion. 
More theoretical work is clearly necessary to 
fully resolve this fundamental problem.

We thank J. Lagerqvist for help in generating the figures in this paper.
We are grateful to Adrian Sutton for invaluable discussions and for the inspiration
he has provided through his deep and long-standing interest in this problem.
Two of us (MD and YCC) acknowledge support from the National Science Foundation 
Grants No. DMR-01-02277 and DMR-01-33075, Carilion Biomedical Institute, and the
Petroleum Research Fund. The calculations reported in this paper were
performed on the beowulf cluster of the Laboratory for Advanced Scientific
Computing and Applications at Virginia Tech.


\begin{thebibliography}{99}
\bibitem{prec} It is worth mentioning at this point that the problem is not just ``academic'', but can have an 
impact in nanoscale electronics. As an example, consider a perfect, defect-free quasi two-dimensional wire and identify two points of the wire in such a way that a straight line connecting these points separates the wire in 
two equivalent regions. If the force is non-conservative, then you can move an atom from one point to the other 
along different paths in the two equivalent regions such that the work done by the current-induced force is different for the different paths. As a consequence, it is in principle possible to break equivalent regions of a defect-free 
wire differently. While this effect might not be important in macroscopic wires, it is relevant in nanoscale 
conductors made of a relatively small number of atoms. A further implication of the problem is 
the following. If current-induced forces are not conservative, then their cross-derivatives 
with respect to ionic positions are not necessarily equal and hence the dynamical response 
matrix for a current-carrying structure is not necessarily symmetric, in which case the 
conventional notion of phonons would be lost in the presence of current.
\bibitem{comment} Analogues of 
Eq.~(\ref{equation}) in basis sets that depend parametrically on the ionic 
positions are available as well.~\cite{tod01}
\bibitem{prec1} For simplicity we may here imagine a jellium wire.
\bibitem{prec12} Both the type of atom and the electron densities 
of the metal electrodes are irrelevant in this numerical demonstration.
\bibitem{prec13} The single-particle wavefunctions are expanded in plane waves in the three spatial directions. 
The number of plane waves in the direction of current flow has been increased with increasing electrode 
separation while keeping the number of plane waves in the perpendicular direction constant. This way, 
we have an estimated error of about 10\% in the current and about 0.5 mRy/a.u. in the force irrespective 
of distance. The maximum jellium-jellium edge separation we can consider to keep the errors within the above values 
is 16 a.u. Other details of the calculations of the current and the force can be found in 
Refs.~\onlinecite{md2,langt}.
\end{thebibliography}
\end{document}